\documentclass[prl,aps,reprint,showpacs,amssymb,superscriptaddress]{revtex4-1}
\usepackage{graphicx}
\usepackage{dcolumn}
\usepackage{bm}
\usepackage[usenames,dvipsnames]{color}

\newcommand{\tg}{\textcolor{Red}}

\newcommand{\tr}{\textcolor{Red}}
\renewcommand{\tr}{}
\renewcommand{\tg}{}

\newcommand{\bk}{\mathbf{k}}
\newcommand{\ua}{\uparrow}
\newcommand{\da}{\downarrow}
\newcommand{\eq}{\begin{equation}}
\newcommand{\eqx}{\end{equation}}
\newcommand{\eqn}{\begin{eqnarray}}
\newcommand{\eqnx}{\end{eqnarray}}
\newcommand{\s}{\sigma}
\newcommand{\veck}{{\bf k}}
\newcommand{\veci}{{\bf i}}
\newcommand{\vecj}{{\bf j}}

\newcommand{\vecl}{{\bf l}}
\begin{document}


\title{Superconductivity in the two-dimensional Hubbard model: Gutzwiller wave function solution}

\author{Jan Kaczmarczyk}
\email{jan.kaczmarczyk@uj.edu.pl}

\author{Jozef Spa\l ek}
\email{ufspalek@if.uj.edu.pl}
\affiliation{Instytut Fizyki im. Mariana Smoluchowskiego, Uniwersytet Jagiello\'{n}ski, Reymonta 4, 30-059 Krak\'ow, Poland}

\author{Tobias Schickling}
\affiliation{Fachbereich Physik, Philipps Universit\"at Marburg,
D-35032 Marburg, Germany}

\author{J\"org B\"unemann}
\email{buenemann@googlemail.com}
\affiliation{Institut f\"ur Physik, BTU Cottbus, D-03013 Cottbus,
Germany}

\date{\today}

\begin{abstract}%
A systematic diagrammatic expansion for Gutzwiller-wave functions (DE-GWF) is formulated and used for the description of superconducting (SC) ground state in the two-dimensional Hubbard model with electron-transfer amplitudes~$t$ (and~$t'$) between nearest (and next-nearest) neighbors. The method is numerically very efficient and allows for \tr{a detailed} analysis of the phase diagram as a function of all relevant parameters ($U$, $\delta$, $t'$) and a determination of the kinetic-energy driven pairing region. SC states appear only for substantial interactions, $U/t \gtrsim 3$, and for not too large hole doping, $\delta \lesssim 0.32$ for $t'=0.25 t$; this upper critical doping value agrees well with experiment for the cuprate high-temperature superconductors. We also obtain other important and novel features of the SC state: (i) the SC gap at the Fermi surface resembles $d_{x^2-y^2}$-wave only around the optimal doping and the corrections to this state are shown to arise from the longer range of the pairing; (ii) the nodal Fermi velocity is almost constant as a function of doping and agrees quantitatively with the experimental results; (iii) the SC transition is driven by the kinetic-energy lowering for low doping and strong interactions.
\end{abstract}



\maketitle

{\sl Introduction.}
High-temperature superconductivity in cuprates is often discussed starting either from the Hubbard model~\cite{Scalapino,Norman2} or from its projected version in the strong-correlation limit, the $t$-$J$ model~\cite{Ogata,Rev_High_Tc}. These models incorporate, in the simplest manner, the strongly correlated nature of the $3d$ electrons due to copper spins in CuO$_2$ planes. The $t$-$J$ model contains real-space operators for their antiferromagnetic coupling explicitly~\cite{Spalek1}. The coupling of the spin degrees is less obvious in Hubbard model unless one introduces antiferromagnetic spin-fluctuations as a pairing mediator from the outset~\cite{Arita,Chubukov}, a model, which can be analyzed reliably only for low values of the Hubbard interaction~$U$. In general, methods are desirable which can treat the Hubbard model for weak to strong correlations, where a possible pairing in momentum space may transform into pairing in real space as a function of $U$. Such evolution of pairing with the increasing interaction strength is particularly interesting in view of the circumstance that iron-pnictide superconductors can be regarded as moderately correlated systems~\cite{Norman,Cvetkovic}.

The Variational Monte Carlo (VMC) method is among the few available numerical many-particle methods which treat the superconducting (SC) state~\cite{Scalapino}. However, it is limited to single-band, small-size systems, containing typically \tr{up to 16 $\times$ 16} lattice sites for the two-dimensional Hubbard model~\cite{Edegger,Lugas,Eichenberger}. Comparable in accuracy (and limitations) is the density-matrix renormalization group approach~\cite{Corboz,Liu}. Lastly, an extensive numerical analysis of the Hubbard model at nonzero temperature and for $t'=0$ has also been carried out within the $2 \times 2$~\cite{Haule} and $8$-site~\cite{Gull} cluster dynamical mean-field theory (DMFT). The normal phase has been investigated on $4\times 4$ cluster~\cite{Jarrell}.

In this work we evaluate the Gutzwiller wave function (GWF) for SC ground state of the two-dimensional Hubbard model.
We extend a recently devised (for the normal state) systematic diagrammatic expansion (DE-GWF), which provides essentially exact results for the GWF up to moderately strong correlations~\cite{Buenemann}. The DE-GWF method has been tested against the exact results in one spatial dimension~\cite{Metzner}, where it removes the spurious Brinkman--Rice metal-insulator transition present in the Gutzwiller approximation and compares favorably with the exact Lieb--Wu solution~\cite{Kurzyk}. In this respect, our approach provides one of the canonical solutions for the SC phase, appearing solely as a result of interparticle correlations.

\begin{figure}[h!]
\centerline{\includegraphics[width=0.9\columnwidth]{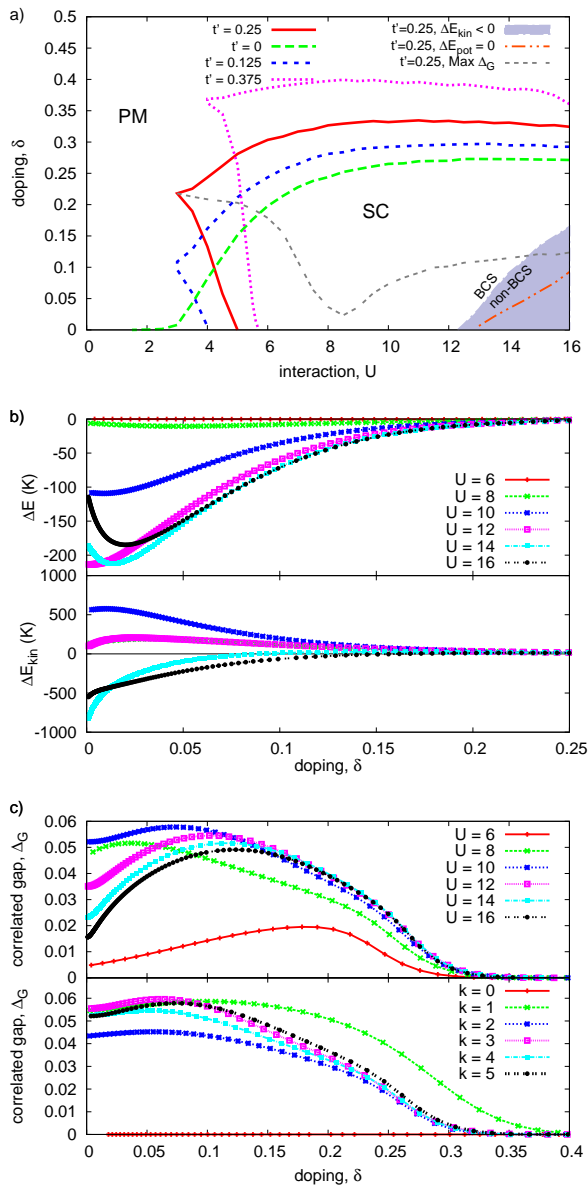}}
\caption{(color online)
(a) Phase diagram comprising paramagnetic (PM) and superconducting (SC) phases as a function of interaction strength~$U$ and doping~$\delta$ for selected values of~$t'$. The gray (dotted) curve marks the optimal doping, the shaded region corresponds to kinetic energy gain in the SC phase (a non-BCS behavior), whereas the curve inside it provides the boundary between the region with positive (below) and negative (above) potential energy change upon condensation.
(b) Top: condensation energy (in Kelvin, for $t = 0.35\, \textrm{eV}$)
as a function of doping for selected values of~$U$. Bottom: the kinetic energy part $\Delta E_{\rm kin}$ of the condensation energy for selected interaction values.
(c) Top: correlated gap as a function of doping;
Bottom: Correlated gap (for $U=10$) in orders $0$--$5$ to which the expansion is carried out.\label{fig:1}}
\end{figure}

\begin{figure}[hb]
\centerline{\includegraphics[width=0.7\columnwidth]{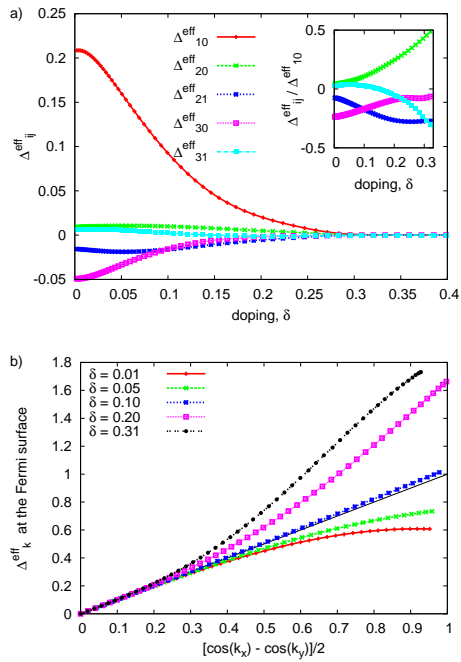}}
\caption{(color online)
(a) Variational gap parameters as a function of doping for $U=10$. Inset: the gap components relative to dominant contribution $\Delta_{10}^{\rm eff}$.
(b) Effective gap in momentum space at the Fermi energy for selected doping values and $U=10$. The black line corresponds to a pure $d_{x^2 - y^2}$ dependence. The gaps are normalized, so that $\Delta_{\bk}^{\rm eff} = 1$ in the anti-nodal direction.
\label{fig:2}}
\end{figure}

\begin{figure}[htb]
\centerline{\includegraphics[width=0.7\columnwidth,angle=270]{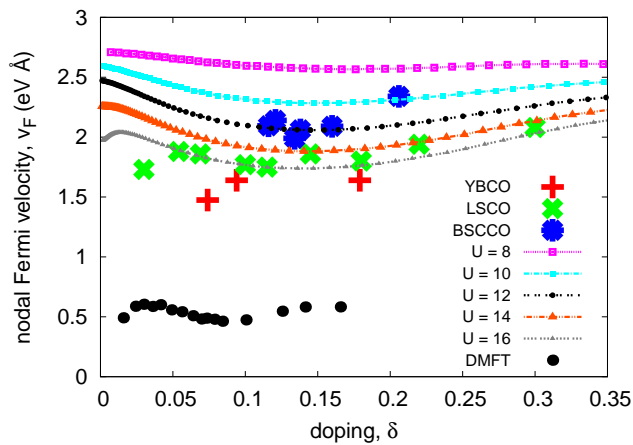}}
\caption{(color online) Universal Fermi velocity in the nodal direction as a function of doping for selected values of $U$. The experimental values are taken from Ref.~\onlinecite{EdeggerPRL} and references therein and have typically an uncertainty of $20\%$. The DMFT results are taken from Ref.~\onlinecite{Civelli1}.
\label{fig:3}}
\end{figure}

\begin{figure}[htb]
\centerline{\includegraphics[width=0.9\columnwidth]{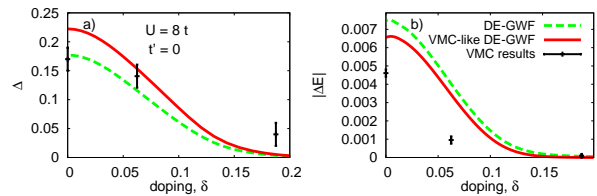}}
\caption{(color online)
Left: Comparison of DE-GWF (lines) for $U=8$ and $t'=0$ with Variational Monte Carlo results (points with error bars; from Ref. \onlinecite{Eichenberger}). (a) Gap parameter $\Delta_{\rm G}$ and (b) condensation energy as a function of doping. The solid lines give the DE-GWF result where the effective single-particle Hamiltonian contains only nearest-neighbor and on-site terms, the dashed lines give the full DE-GWF result.
\label{fig:4}}
\end{figure}

Our method is numerically very efficient so that we can determine \tr{a detailed} ground-state phase diagram of the Hubbard model, with normal (paramagnetic, PM) and SC phases as a function of the Hubbard interaction~$U$, the hole doping~$\delta$, and $t'$.
One principal advantage of our approach is the ability to account, for nonzero pairing amplitudes beyond the nearest neighbors (n.n.). In the following we study the doping dependence (Fig.~\ref{fig:1}c) and $\bk$-dependence (Fig.~\ref{fig:2}) of the SC gap obtaining deviations from the $d_{x^2 - y^2}$-wave gap symmetry.
We investigate the kinetic energy gain upon the condensation (Fig.~\ref{fig:1}b) and the nodal Fermi velocity (Fig.~\ref{fig:3}) to show that the present approach reproduces the principal experimental findings. \tg{We also compare our results with those of VMC (Fig.~\ref{fig:4}).}

{\sl Method.}
The main features of the DE-GWF method for the PM state have been provided in
Ref.~\onlinecite{Buenemann}. Here, we summarize the essential steps
and subsequently generalize the approach to the description of
SC ground states. We start from the Hubbard Hamiltonian on $L$~sites
of a square lattice
\begin{equation}
\hat{H}=\hat{H}_0 + U\sum_{\veci} \hat{d}_{\veci} \,,
\hat{H}_0=\sum_{\veci,\vecj,\sigma}t_{\veci\vecj}\hat{c}_{\veci,\sigma}^{\dagger}
\hat{c}_{\vecj,\sigma}^{\phantom{\dagger}} \, ,
\hat{d}_{\veci}\equiv\hat{n}_{\veci,\uparrow}\hat{n}_{\veci,\downarrow} \, ,
\label{eq:1}
\end{equation}
where $\veci = (i_1, i_2)$ is the two-dimensional site-index,
$t_{\veci \vecj} = -t$ and $t'$ are
the hopping integrals for nearest and for next-nearest neighbors,
respectively, and $\s = \ua, \da$ is
the spin quantum number. The Gutzwiller wave function~\cite{Gutzwiller}
for the correlated state has the form
\begin{equation}
|\Psi_{\rm G}\rangle = \hat{P} |\Psi_0\rangle =
\prod\nolimits_{\veci}\hat{P}_{\veci}|\Psi_0\rangle \; ,
\label{eq:1.2}
\end{equation}
where $|\Psi_0\rangle$ is a single-particle
product state (Slater determinant) to be defined later.
We define the local Gutzwiller correlator as
\begin{eqnarray}
\hat{P}_{\veci} &\equiv&
\sum_{\Gamma}\lambda_{\Gamma} |\Gamma \rangle_{\veci\,\veci}\! \langle
\Gamma | \; , \label{eq:proj1} \\
\hat{P}^2_{\veci} &\equiv&
1+x\hat{d}_{\veci}^{\rm HF} \; . \label{eq:proj2}
\end{eqnarray}
Eq.~(\ref{eq:proj1}) presents a general form of the correlator
with variational parameters $\lambda_\Gamma
\in \left\{ \lambda_\emptyset, \lambda_{1\ua}, \lambda_{1\da},
\lambda_d \right\}$, which describe the occupation probabilities
of the four possible local states $ \{ |\Gamma\rangle_{\veci} \} \equiv
\left\{|\emptyset\rangle_{\veci}, |\uparrow\rangle_{\veci},
|\downarrow\rangle_{\veci}, |\uparrow\downarrow\rangle_{\veci}\right\}$.
In Eq.~(\ref{eq:proj2}), a particularly useful
form of the local correlator is given, where
the Hartree--Fock operators are defined by
$\hat{d}_{\veci}^{\rm HF} \equiv
\hat{n}^{\rm HF}_{\veci,\uparrow}\hat{n}^{\rm HF}_{\veci,\downarrow}$ and
$\hat{n}^{\rm HF}_{\veci,\sigma}\equiv\hat{n}_{\veci,\sigma}-n_0 $
with $n_0 = \langle \Psi_0| \hat{n}_{\veci,\sigma} |\Psi_0 \rangle$.
This form of $\hat{P}^2_{\veci}$ decisively simplifies the
calculations by eliminating the `Hartree bubbles'~\cite{Gebhard,Buenemann}.

We calculate all required expectation values diagrammatically as a power series in $x$: the norm, $\langle\Psi_{\rm G}|\Psi_{\rm G} \rangle$, the double occupancy $\langle\Psi_{\rm G}|\hat{d}_{\veci}|\Psi_{\rm G} \rangle \equiv \langle \hat{d}_{\veci}\rangle_G$,
and the hopping term $\langle\hat{c}^{\dagger}_{\veci,\sigma} \hat{c}_{\vecj,\sigma}^{\phantom{\dagger}} \rangle_G$, see Refs.~\onlinecite{Buenemann,SM} for details. Here, we discuss the new features appearing in the presence of SC pairing.

First, apart from the `normal' lines, as represented by $P_{\vecl,\vecl'} \equiv P^{\sigma}_{\vecl,\vecl'} \equiv \langle \Psi_0 | \hat{c}^{\dagger}_{\vecl,\sigma} \hat{c}_{\vecl',\sigma}^{\phantom{\dagger}}| \Psi_0 \rangle -\delta_{\vecl,\vecl'}n_0$, we also have to take into account the anomalous (SC) lines $S_{\vecl,\vecl'} \equiv \langle \Psi_0 | \hat{c}^{\dagger}_{\vecl,\uparrow} \hat{c}^{\dagger}_{\vecl',\downarrow} | \Psi_0 \rangle$, what leads to much more involved computations as there are up to 1000 times more SC diagrams than PM diagrams in the fifth order. Note that we consider only the $d$-wave spin-singlet SC order without a local pairing, i.e., with $S_{\vecl,\vecl}\equiv 0$.

Second, since the correlated number of
particles, $n_{\rm G} \equiv \langle \hat{n}_{\veci,\sigma} \rangle_{\rm G}$
and its non-correlated correspondent $n_0$ may differ in the
SC phase, the minimization procedure is different. Namely,
we minimize the generalized grand-canonical potential $\mathcal{F} =
\langle \hat{H} \rangle_{\rm G} - 2 \mu_{\rm G} n_{\rm G} L$
instead of minimizing the ground-state energy
$E_{\rm G} \equiv \langle \hat{H} \rangle_{\rm G}$.

Third, the minimization procedure leads to an effective
single-particle Hamiltonian which, in the present situation, contains
also the SC pairing contribution,
\begin{eqnarray}
\hat{H}_0^{\rm eff} &=&
\sum_{\veci,\vecj,\sigma}t^{\rm eff}_{\veci,\vecj}
\hat{c}_{\veci,\sigma}^{\dagger}\hat{c}_{\vecj,\sigma}^{\phantom{\dagger}}
+ \sum_{\veci,\vecj} \bigl( \Delta^{\rm eff}_{\veci,\vecj}
\hat{c}_{\veci,\ua}^{\dagger}\hat{c}_{\vecj,\da}^{\dagger} +
{\rm h.c.} \bigr)\; , \label{eq:iou1}\\
t^{\rm eff}_{\veci,\vecj} &=&
\frac{\partial \mathcal{F} (|\Psi_0\rangle,x)}{\partial P_{\veci,\vecj}}
\;, \quad \Delta^{\rm eff}_{\veci,\vecj} =
\frac{\partial \mathcal{F} (|\Psi_0\rangle,x)}{\partial S_{\veci,\vecj}} \;.
\label{eq:iouF}
\end{eqnarray}
{}From $\hat{H}_0^{\rm eff}$
we can deduce the quasi-particle dispersion $\epsilon^{\rm eff}(\veck)=
(1/L)\sum_{\veci,\vecj}t^{\rm eff}_{\veci,\vecj}\exp[i \bk \cdot (\veci-\vecj)]$,
and the quasi-particle gap function $\Delta^{\rm eff}(\veck)
=(1/L)\sum_{\veci,\vecj}\Delta^{\rm eff}_{\veci,\vecj}\exp[i \bk \cdot (\veci-\vecj)]$.
The latter must be distinguished from the correlated gap, defined by
$ \Delta_{\rm G} \equiv
\langle c_{\veci \ua}^\dagger c_{\vecj \da}^\dagger \rangle_{\rm G} $
for n.n. $\langle \veci, \vecj \rangle$, see Ref.~\onlinecite{SM}
for an explicit analytical expression. The Hamiltonian $\hat{H}_0^{\rm eff}$ also defines $|\Psi_0\rangle$ which is its ground state.

{\sl Results.}
If not stated otherwise, we present the results to the fifth order of the expansion for the parameter value $t' = 0.25 t$, with $t = 1$ as our unit of energy. Moreover, we take into account only those lines $P_{\veci, \vecj} \equiv P_{0, (\veci-\vecj)} \equiv P_{XY}$ (with $X = i_1 - j_1$, $Y = i_2-j_2$) which fulfill $X^2 + Y^2 \leq 10$. The same condition applies for $S_{\veci, \vecj}$, $t^{\rm eff}_{\veci, \vecj}$, and $\Delta^{\rm eff}_{\veci, \vecj}$. We have checked that this truncation in real space does not influence the results qualitatively in the parameter regime discussed in this work.
Note that the complete phase diagram is calculated within a few days on a modern PC.

Fig.~\ref{fig:1} summarizes the ground-state characteristics of the SC phase \tr{(defined as that with $\Delta_{10}^{\rm eff} > 10^{-4}$)}. As can be seen from Fig.~\ref{fig:1}a, the SC region expands with increasing $t'$ towards higher doping values \cite{Shih,Yokoyama2}. For fixed $t'$, the critical value $\delta_c$ above which the SC state disappears is fairly independent of~$U$ (for $U \gtrsim 8$) and the universal value $\delta_c \approx 0.32$ (for $t'=0.25$) is in good agreement with experimental data for virtually all single-plane cuprates and with recent sophisticated renormalized mean-field theory (RMFT) calculations for the $t$-$J$ model~\cite{Jedrak2}. The reentrant behavior of the SC phase as a function of doping is associated with the dome-like SC (cf. the $U = 6$ curve in (c)). \tr{The onset of SC phase requires a minimal on-site interaction $U>3$ even for the optimal doping. There may still be a tendency towards SC below $U = 3$ and above $\delta_c = 0.32$: we see an exponential tail of the gap and the condensation energy in this regime, similarly as in Ref.~\onlinecite{Yokoyama2}.}

The condensation energy $\Delta E \equiv E_{\rm G}^{\rm (SC)} - E_{\rm G}^{\rm (PM)}$ shown in Fig.~\ref{fig:1}b is measured in Kelvin (for $t = 0.35\, \textrm{eV}$). It shows that \tr{our method} provides an energy gain in the proper range of the critical temperature for the cuprates. The corresponding kinetic energy change $\Delta E_{kin}$ in Fig.~\ref{fig:1}b (bottom) proves that the superconductivity is kinetic-energy driven \cite{Deutscher,Gedik,Giannetti,Carbone,Yokoyama1,Yanase,Yokoyama2,Haule,Gull,Singh,Hirsch2,Hirsch} for the cases of low doping and the large interaction values \tr{$U \gtrsim 12$, in agreement with Refs.~\onlinecite{Yokoyama1,Yokoyama2} analyzing more sophisticated wave functions}. This region is marked in the phase diagram (cf. Fig.~\ref{fig:1}a) as the shaded area. For $U=14 \div 16$, the doping at which superconductivity becomes kinetic-energy driven coincides with the optimal doping, in agreement with the experimental results for the cuprates \cite{Deutscher,Gedik,Giannetti,Carbone}. This phenomenon has also been studied theoretically within \tr{the VMC \cite{Yokoyama1,Yanase,Yokoyama2} method}, as well as within the cluster DMFT for the $t$-$J$ \cite{Haule} and, very recently, the Hubbard \cite{Gull} models. \tr{The DMFT} studies are limited to nonzero temperature (e.g. $\beta = 60/t$ in Ref.~\onlinecite{Singh}) and $t'=0$ what has been pointed out~\cite{Carbone} as a possible source of a quantitative disagreement with experimental results. The validity of the $t$-$J$ model for such analysis has been disputed in view of the virial theorem violation \cite{Singh}.

In Fig.~\ref{fig:1}c the correlated gap shows a dome-like structure as a function of doping for $U \gtrsim 10$. The maximal value for the correlated gap is achieved for $U \gtrsim 10$ near doping $\delta\approx 0.1$ \cite{Yokoyama2}. If one takes $\Delta_{\rm G}$ as a measure of the superconductivity strength, one can conclude that moderate to strong interactions and not too small dopings are optimal for superconductivity as is also observed for cuprate superconductors. Since the results for $\Delta_G$ obtained in the fourth and the fifth orders (see Fig.~\ref{fig:1}c, bottom) do not differ remarkably for the investigated parameter range, we may say that our method provides very accurately the ground-state properties of GWF for moderate to strong correlations. Note also that the zeroth-order calculations, which can be viewed as a sophisticated renormalized mean-field theory (RMFT) calculations, do not yield a stable SC state.

In the Gutzwiller approach, the structure of the gap function in the effective single-particle Hamiltonian~$\hat{H}_0^{\rm eff}$
is optimized variationally. The effective dispersion relation of the Hamiltonian~(\ref{eq:iou1}) defines the quasi-particle spectrum and is thus related~\cite{EdeggerPRL} to the quasi-particle peaks observed in photoemission experiments. Fig.~\ref{fig:2}a shows the effective components of the quasi-particle ($d$-wave symmetry) gap function $\Delta_{\veci, \vecj}^{\rm eff} \equiv \Delta^{\rm eff}_{XY} = - \Delta^{\rm eff}_{YX}$ (with $X=i_1-j_1$ and $Y=i_2-j_2$) as a function of doping. The dominant component is the n.n. contribution $\Delta_{10}^{\rm eff}$, so that the gap has mainly $d_{x^2 - y^2}$ dependence. However, the other components, particularly $\Delta_{30}^{\rm eff}$ and $\Delta_{21}^{\rm eff}$ lead to a noticeable deformation of the gap function away from the optimal doping, $\delta \sim 0.1$, as shown in Fig.~\ref{fig:2}b which displays the effective gap in reciprocal space across the Fermi surface. The deviations from the $d_{x^{2}-y^{2}}$-dependence are most prominent in the anti-nodal direction. Such deviations have been observed in high-Tc superconductors~\cite{Mesot,Yoshida,Lee,Vishik,Vishik3} \tr{and investigated theoretically within VMC~\cite{Watanabe} (without inclusion of effective hoppings beyond third-nearest neighbors). Inclusion of \tg{the longer-range effective parameters} is usually omitted in VMC probably because of the computational cost. }
Our results do not necessarily reflect the physics of this phenomenon in cuprate superconductors where the deviation may be caused by two energy scales corresponding to a two-gap structure~\cite{Yoshida}. Note that in the overdoped regime ($\delta \geq 0.2$) the gap components become of comparable magnitude \tg{(cf. inset in Fig.~\ref{fig:2}a)}. This may be interpreted as a gradual evolution from real-space pairing for the optimal doping to momentum-space pairing close to the upper critical concentration $\delta_c \approx 0.32$.

One of the most important physical characteristics of the cuprates is the universal nodal Fermi velocity $v_F$~\cite{Zhou} (i.e., $v_F$ is independent of $\delta$). This quantity, defined as $v_F = \nabla_\bk \epsilon^{\rm eff}(\bk)|_{\epsilon^{\rm eff}(\bk) = 0}$, is exhibited in Fig.~\ref{fig:3} and the trend agrees very well with the experimental results (we assume the lattice constant $a=4\, {\rm \AA}$ and $t=0.35\, {\rm eV}$). We also show the DMFT results for the Hubbard model \cite{Civelli1,Civelli2} in the physical units (assuming the same values of $a$ and $t$). RMFT does not reproduce such behavior \cite{Jedrak2,EdeggerPRL} due to lack of momentum-space differentiation \cite{Civelli2} (i.e. band renormalization factors $q_\sigma$ are independent of $\bk$), whereas the VMC results were obtained (to the best of our knowledge) only for the $t$-$J$ model \cite{Paramekanti,Yunoki}. Therefore, our results provide the first quantitative agreement for the Hubbard model. \tr{Note however, that recently the Fermi velocity for the underdoped samples has shown a doping dependence \cite{Vishik2}. The result of Ref. \onlinecite{Vishik2} is that the velocity has the two components: one near the Fermi surface which is doping dependent and the velocity slightly below the Fermi surface which is doping independent. We believe that a purely electronic model should provide only a doping-independent nodal Fermi velocity \tg{(cf. also Ref.~\onlinecite{Johnston}).}}

In Fig.~\ref{fig:4} we compare the results of our DE-GWF with VMC results of Ref.~\onlinecite{Eichenberger}, obtained for $U=8$ and $t'=0$. The `VMC-like' DE-GWF results were obtained in the fifth order by setting the effective parameters $t_{\veci,\vecj}^{\rm eff}$ and $\Delta_{\veci,\vecj}^{\rm eff}$ to zero beyond n.n. Moreover, we use $t^{\rm eff}_{10}\equiv -t$, and $\Delta^{\rm eff}_{10} \equiv \Delta$, \tg{as well as $t^{\rm eff}_{00}$} as our remaining variational parameters~\cite{SM}. The data `DE-GWF' are the result of the full fifth-order expansion.

The VMC results~\cite{Eichenberger} and the DE-GWF VMC-like results are close to each other near the half-filling, with quantitative differences away from half filling. \tg{The sources of these discrepancies are approximations of both methods. First, in VMC calculations,} an $8\times8$ lattice is used which may be too small to emulate the infinite lattice used in DE-GWF. \tr{This can be seen explicitly from Ref.~\onlinecite{EichenbergerPHD} (cf. Fig. 3.21), where the extrapolation of the gap value in the thermodynamic limit is shown. The nonzero gap obtained \tg{at $\delta \approx 0.19$} by VMC for the $8 \times 8$ system extrapolates to zero gap in the thermodynamic limit obtained from finite-size scaling (which agrees with our result in Fig. 4.).} \tg{Second, in our method we perform the expansion up to the 5th order and we use the $|\Psi_0 \rangle$ lines up to 7th neighbors.}

\tg{Differences between the `full' and `VMC-like' DE-GWF curves show that neglecting the longer range effective parameters can lead to the decrease of the condensation energy by $11 \%$ and the increase of the principal gap component ($d_{x^2-y^2}$-wave) by $26 \%$.}


\tg{The DE-GWF method in the present formulation is taylor-made for the Gutzwiller Wave Function. More general wave functions have been shown to improve the energy (e.g. wavefunctions with the doublon-holon correlation \cite{Yokoyama1,Yokoyama2} or Baeriswyl wavefunctions \cite{Baeriswyl,Hetenyi,Eichenberger}). Investigation of the possibility of extension of the DE-GWF method in this direction is planned.}

{\sl Summary.}
We have formulated an efficient diagrammatic evaluation of the Gutzwiller-correlated wave function and have carried out our DE-GWF to the fifth order for the superconducting (SC) ground state. Our approach works in the thermodynamic limit and for general single-particle states $|\Psi_0\rangle$ (with \tr{the effective pairing and hopping} \tr{taken up to 7th neighbors in the present study}), whereby we overcome the limitations of the Variational Monte Carlo method. The DE-GWF method allows for detailed investigation (as a function of all relevant parameters) of fundamental phenomena for the cuprates: the universal nodal Fermi velocity, the kinetic-energy driven (non-BCS) superconductivity, and the deviations from the $d_{x^2-y^2}$ gap symmetry. We obtain agreement with the experimental results (in some cases better than for any other method). We also provide a comprehensive phase diagram of superconductivity in the Hubbard model comprising the non-BCS regime of pairing.

A competition or coexistence of SC with antiferromagnetic, and/or Pomeranchuk phases, as well as the extension to multi-band systems is cumbersome but feasible, and should be investigated separately.

\begin{acknowledgments}
The authors are very grateful to Florian Gebhard for discussions and critical reading of the manuscript. We are also grateful to Jakub J\c{e}drak and Dirk van der Marel for discussions. The work was supported in part by the Foundation for Polish Science (FNP) under the `TEAM' program, as well as by the project `MAESTRO' from National Science Centre (NCN), No.\ DEC-2012/04/A/ST3/003420.
\end{acknowledgments}

\bibliography{DiagExp}

\end{document}